\documentclass{pos}

\title{Search for a Low Mass Standard Model Higgs Boson at D\O\ in ppbar Collisions at $\sqrt{s}=1.96 TeV$}

\ShortTitle{Search for a Low Mass Standard Model Higgs Boson at D\O\ in ppbar Collisions at $\sqrt{s}=1.96 TeV$}

\author{\speaker{Murilo Rangel}\thanks{On behalf the D\O\ collaboration}\\
        Laboratoire de l'Acc\'el\'erateur Lin\'eaire\\
        E-mail: \email{rangel@fnal.gov}}

\abstract{We present combined searches for the Low Mass Standard Model Higgs boson
at a center-of-mass energy of $\sqrt{s}=$1.96~TeV, using up to 5~fb$^{-1}$ of
data collected with the D0 detector at the Fermilab Tevatron collider. The
major contributing processes are associated production ($WH\to l\nu bb$,
$ZH\to\nu\nu bb$, $ZH\to ll bb$). The significant improvements across the full mass
range resulting from the larger data sets and improved analyses as well as future
prospects are discussed.}

\FullConference{XXth Hadron Collider Physics Symposium \\
		 November 16 -- 20, 2009 \\
		 Evian, France}

\begin{document}

The standard model (SM) of particle physics is a succefull theory despite
from the fact that the electroweak symmetry breaking mechanism is still
unknown. The simplest proposed mechanism involves the introduction of a
complex doublet of scalar fields that generate the masses of elementary
particles via their mutual interactions. After accounting for longitudinal
polarizations for the electroweak bosons, this so-called Higgs mechanism 
also gives rise to a single scalar boson (SM Higgs) with an unpredicted mass.

We combined results of different direct searches for Low Mass SM Higgs
(mass less than 135 GeV) recorded by the D\O\ experiment \cite{combo}.
The analyses serch for a signal of Higgs bosons produced in association with vector
bosons ($q\bar{q} \to W/ZH$), through gluon-gluon fusion (GGF) ($gg \to H$), 
through vector boson fusion (VBF) ($q\bar{q} \to q'\bar{q'} H$),
and in association with top quarks ($t\bar{t} \to t\bar{t}H$).
The analyses utilize data corresponding to integrated luminosities
ranging from 2.1 to 5.4 fb$^{-1}$, collected during the period 2002-2009.
The Higgs boson decay modes studied are $H\to b\bar{b}$, $H\to W\bar{W}$, 
$H\to \tau \bar{\tau}$ and $H\to \gamma \gamma$.
The analyses were designed to be mutually exclusive after analysis selections.

Since the most recent D\O\ SM combined Higgs boson search results \cite{oldcomb},
from the analyses sensitive to Low Mass Higgs, the $ZH \to \nu \bar{\nu} b \bar{b}$
was updated \cite{nunubb}.

The main Higgs decay at Low Mass is $H\to b\bar{b}$, therefore,
an algorithm for identifying jets consistent with
the decay of a heavy-flavor quark is applied to each jet (b-tagging).
Several kinematic variables sensitive to displaced jet vertices and jet tracks with 
large transverse impact parameters relative to the hard-scatter vertices are combined
in a neural network (NN) discriminant trained to identify heavy-flavor quark decays and
reject jets arising from light-flavor quarks or gluons \cite{btagging}.

We combine results using the $CL_{s}$ method with a negative log-likelihood ratio (LLR) test statistic.
The value of $CL_{s}$ is defined as $CL_{s}$ = $CL_{s+b}$ /$CL_{b}$ where $CL_{s+b}$ and $CL_b$ are the confidence levels
for the signal-plus-background hypothesis and the background-only hypothesis, respectively.
The confidence levels are evaluated by integrating
corresponding LLR distributions populated by simulating outcomes via Poisson statistics. Separate channels and bins
are combined by summing LLR values over all bins and channels.
Systematics are treated as Gaussian uncertainties on the expected number of signal and background events, not the
outcomes of the limit calculations.
The $CL_{s}$ approach used in this combination uses binned final-variable
distributions rather than a single-bin (fully integrated) value for each contributing analysis.
The exclusion criteria are determined by increasing the signal cross section until $CL_{s} = 1 - \alpha$,
which defines a signal cross section excluded at 95\% confidence level for $\alpha$ = 0.95.

In order to minimize the degrading effects of systematics on the search sensitivity, the individual background contributions
are fitted to the data observation by maximizing a likelihood function for each hypothesis \cite{collie}. The likelihood is a
joint Poisson probability over the number of bins in the calculation and is a function of the nuisance parameters in
the system and their associated uncertainties, which are given an additional Gaussian constraint associated with their
prior predictions. The maximization of the likelihood function is performed over the nuisance parameters. A fit is
performed to both the background-only (b) and signal-plus-background (s+b) hypothesis separately for each Poisson
MC trial.

We derive limits on SM Higgs boson production cross section times the SM Higgs
branch ratio.
The limits are derived at 95\% C.L. To facilitate model transparency and to accommodate analyses with different
degrees of sensitivity, we present our results in terms of the ratio of 95\% C.L. upper cross section limits to the SM
predicted cross section as a function of Higgs boson mass. The SM prediction for Higgs boson production would
therefore be considered excluded at 95\% C.L. when this limit ratio falls below unity.

\begin{figure}[ht]
\begin{center}
\includegraphics[width=0.49\textwidth]{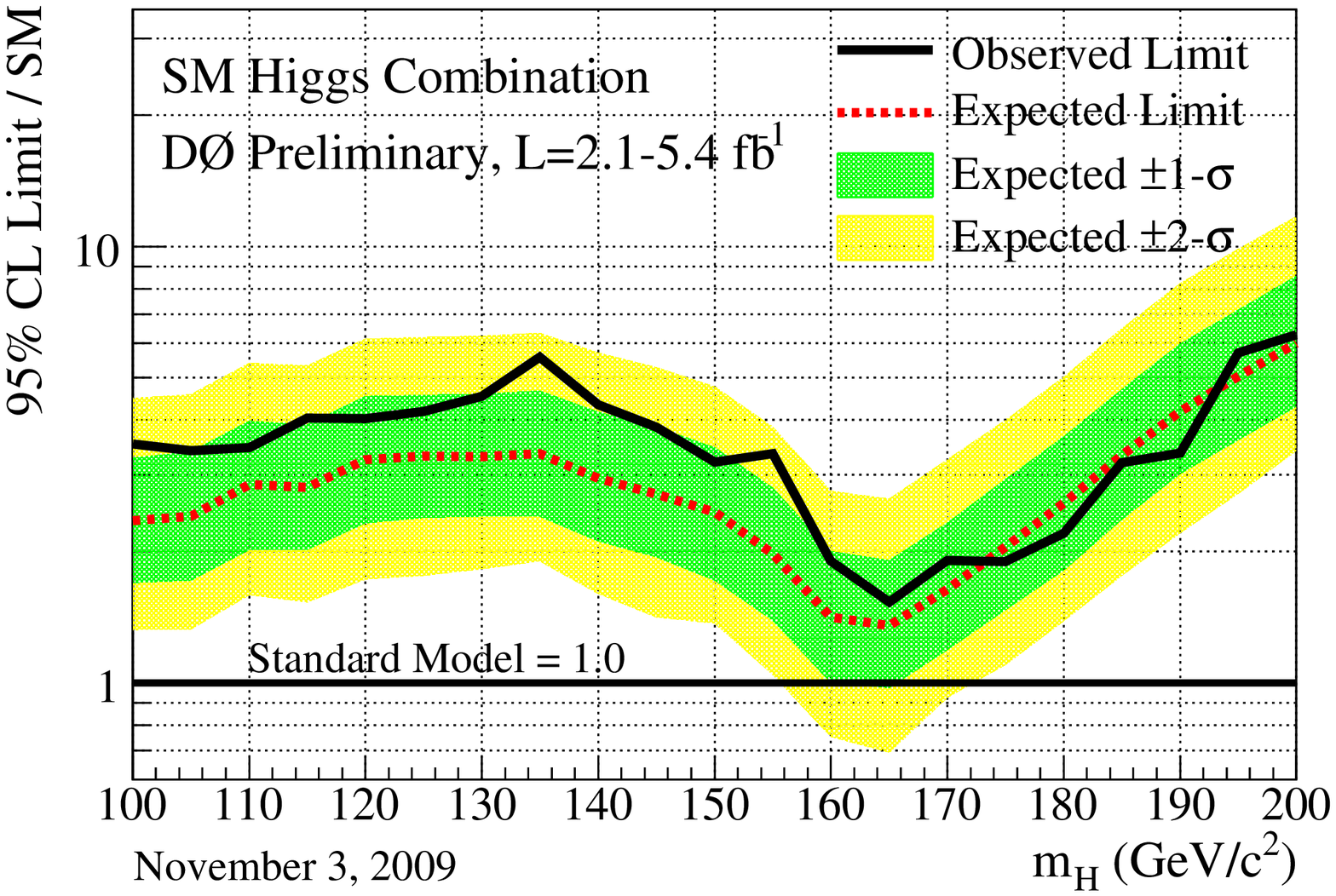}
\includegraphics[width=0.49\textwidth]{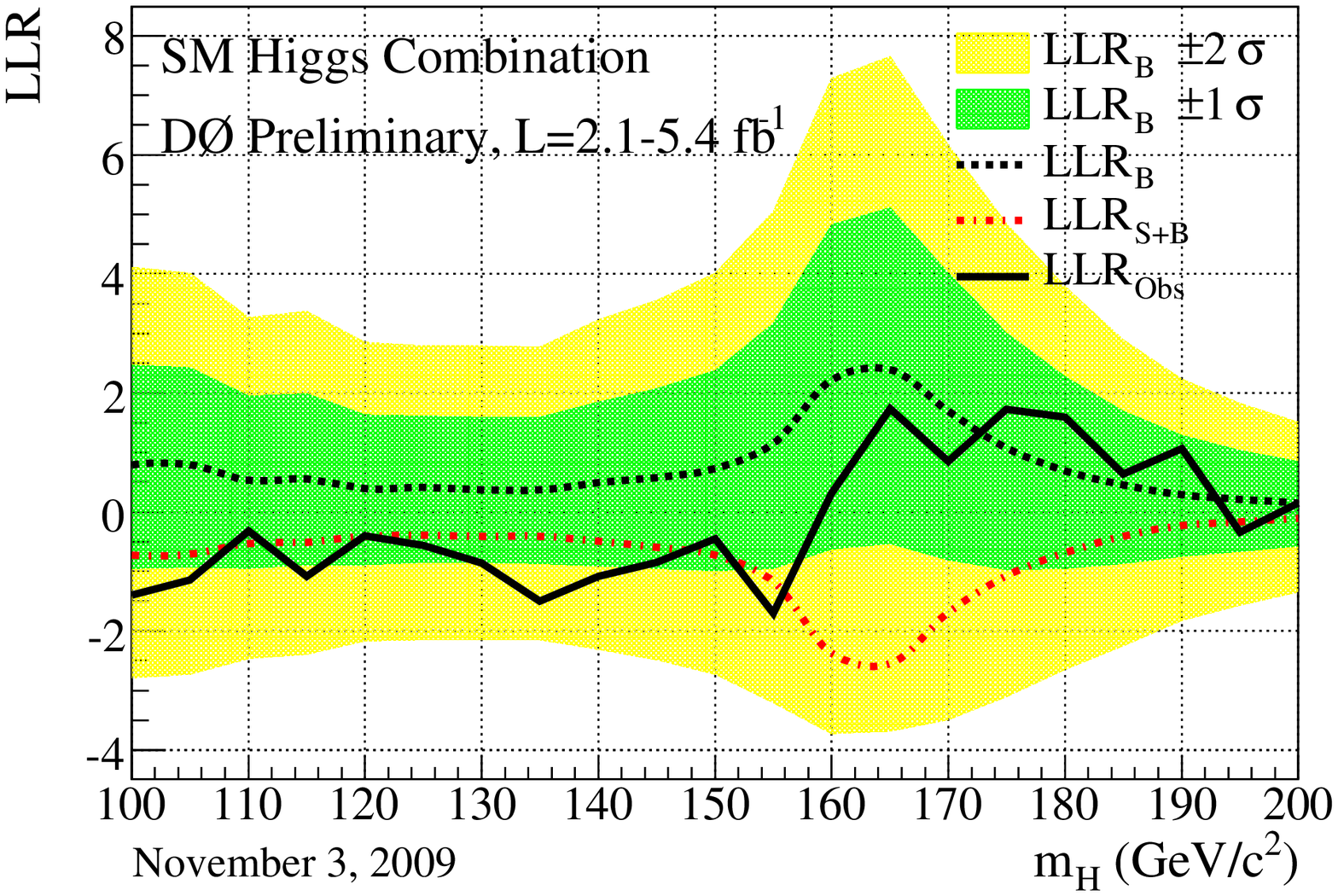}
\caption{
Left: Expected (median) and observed 95\% C.L. cross section upper limit ratios for the combined
Higgs analyses over the 100 $\leq$ $m_H$ $\leq$ 200 GeV/$c^{2}$ mass range.
Right: Log-likelihood ratio distribution for the combined analyses over the 
100 $\leq$ $m_H$ $\leq$ 200 GeV/$c^{2}$ mass range. 
\label{fig:limit}}
\end{center}
\end{figure}

The expected and observed 95\% C.L. cross section limit ratios to the SM cross sections for all
analyses combined over the probed mass region (100 $\leq$ $m_H$ $\leq$ 135 GeV/$c^2$ ) are presented in
Table \ref{fig:limit}.
These limits and the LLR distributions for the full combination are shown in Fig. \ref{fig:limit}.

\begin{table}[!ht]
\begin{center}
\caption{Combined 95\% C.L. limits for SM Higgs boson production. The limits are
reported in units of the SM production cross section times branching fraction.}
\label{tab:trigger}
\begin{tabular}[t]{c c c c c c c c c}
\hline
$m_H$ (GeV/$c^{2}$)&100&105&110&115&120&125&130&135\\
\hline
Expected:&2.35&2.40&2.85&2.80&3.25&3.31&3.30&3.35\\
Observed:&3.53&3.40&3.47&4.05&4.03&4.19&4.53&5.58\\
\hline
\hline
\end{tabular}
\end{center}
\end{table}

  We have presented upper limits on standard model Higgs boson production derived from 60 Higgs search analyses
including data corresponding to 2.1-5.4 fb$^{-1}$. We have combined these analyses and form new limits
more sensitive than each individual limit. The observed (expected) 95\% C.L. upper limit ratios to the SM Higgs
boson production cross sections are 4.0 (2.8) at $m_H$ = 115 GeV/c2

\end{document}